\documentclass[preprint,nofootinbib,APS]{revtex4}
\usepackage{amsmath,amssymb}
\usepackage{graphicx}
\usepackage{psfrag}

\begin{document}

\title{
Finite Energy Monopoles in Non-Abelian Gauge Theories on Odd-dimensional Spaces
}

\author{Hironobu Kihara}
\affiliation{ Korea Institute for Advanced Study\\
207-43 Cheongnyangni 2-dong, Dongdaemun-gu, Seoul 130-722, Republic of Korea
}

\preprint{KIAS-P08064}
\date{\today}

\begin{abstract}
In higher dimensional gauge theory, we need energies with higher power terms of field strength  in order to realize point-wise monopoles. 
We consider new models with higher power terms of field strength and extraordinary kinetic term of scalar field. 
Monopole charges are computed as integrals over spheres and they are related to mapping class degree. 
Hedge-Hog solutions are investigated in these models.
Every differential equation for these solutions is Abel's differential equation. 
A condition for existence of finite energy solution is shown. 
Spaces of 1-jets of these equations are defined as sets of zeros of polynomials. 
Those spaces can be interpreted as singular quartic surfaces in three-dimensional complex projective space. 
\end{abstract}

\maketitle

\section{Introduction}
Monopoles and their generalization  have been studied by many people \cite{Dirac:1948um}. 
Especially topological property of monopoles was studied in \cite{Arafune:1974uy}. 
Tchrakian continues studying the higher-dimensional generalization intensively \cite{Tchrakian:1978sf}. 
The generalization of Dirac-Yang monopole has been considered by \cite{Ito:1984wu}. 
Recently we studied the Tchrakian type-monopole in five-dimensional space \cite{Kihara:2004yz}.

In this article, we would like to consider monopoles in spaces whose dimension is $D=2n+1$ with arbitrary positive integer $n$.  
Our monopoles are realized in Spin($2n+1$) gauge theories, where  
the group Spin($N$) is the universal covering group of the orthogonal group SO($N$). 
In other words, the corresponding Lie algebras of Spin($N$)  and SO($N$) are isomorphic and Spin($N$)  is simply connected. 
We can embed groups SU($N$) into SO($2N$) which is a subgroup of SO($2N+1$) and compact simple Lie groups are embedded  into unitary groups. 
We embed the adjoint representation and vector representation into the corresponding Clifford algebra.  
We will discuss topological currents in these models. 
The topological charges obtained from these currents are proportional to the Brower mapping class degree of spheres. 
We will consider Hedge-Hog solutions and we obtain a condition on the existence of finite energy solutions in our models. 
Every differential equation for Hedge-Hog solutions becomes Abel's differential equations. 
The spaces of 1-jets in term of Arnold \cite{Arnold:1983} are parts  of projective quartic surfaces. 
They are singular and are not K3 surfaces. 

This article is constructed as follows. 
In section 2, we will define our models which are labeled by an integer $p$ and will study topological currents. 
In section 3, we will consider solutions with Hedge-Hog ansatz and derive the Abel's differential equations. 
The case $n=p$ is special and we will divide our discussion into two parts. 
In section 4, we will study the spaces of 1-jets and the property of surfaces. 
In section 5, we will show the condition of existence of solutions. 
The consideration shows that in the case of $p < n$, $p$ has upper bound. 
Finally in section 6, we will summarize our result. 

In the remaining part of this introduction we will review the basic of Clifford algebra and differential forms, which we need in this article. 
 let us explain our notation on Clifford algebra with respect to Euclidean metric and differential forms.
\subsection{Clifford Algebra} 
Let us use  a Clifford algebra with respect to the Euclidean metric on ${\mathbb R}^{2n+1}$, which is generated by gamma matrices $\gamma_a$, $(a=1,2, \cdots , 2n+1$). 
The gamma matrices $\gamma_a$ satisfy the following anti-commutation relation: $\{ \gamma_a , \gamma_b \} = 2 \delta_{ab} $. 
Suppose that $\gamma_a$ are Hermitian. 
 $\gamma_{a(1)}\gamma_{a(2)} \cdots \gamma_{a(\ell)}$ can be decomposed by the anti-commutation relation,  
where $1 \leq a(i) \leq D$ for all $i$. 
Let us use the following notation. 
\begin{align}
\gamma_{a(1)a(2) \cdots a(\ell)} &:= \frac{1}{\ell !} \sum_{\sigma \in {\mathfrak S}_{\ell}} {\rm sgn}(\sigma) \gamma_{a(\sigma(1) )} \gamma_{a(\sigma(2) )}  \cdots \gamma_{a(\sigma(\ell) )}  . 
\end{align}
where $ {\mathfrak S}_{\ell}$ is the symmetric group which consists of permutations of $\{ 1,2, \cdots ,\ell \}$. 
For a permutation $\sigma$, ${\rm sgn}(\sigma)$ is denoted the sign of $\sigma$. 
Multiplications of $\gamma_b $ and $\gamma_{a(1)a(2) \cdots a(\ell)}$ become
\begin{align}
\gamma_b \gamma_{a(1)a(2) \cdots a(\ell)} &= \gamma_{ba(1)a(2) \cdots a(\ell)} + \sum_{i=1}^{\ell} (-1)^{i-1} \delta_{b, a(i)} ~\gamma_{a(1)a(2) \cdots \check{a(i)} \cdots a(\ell)} ,\cr
 \gamma_{a(1)a(2) \cdots a(\ell)} \gamma_b&= \gamma_{a(1)a(2) \cdots a(\ell)b} + \sum_{i=1}^{\ell} (-1)^{l-i} \delta_{b, a(i)} ~\gamma_{a(1)a(2) \cdots \check{a(i)} \cdots a(\ell)} .
\end{align}
And we obtain graded commutation relation
\begin{align}
 \gamma_{a(1)a(2) \cdots a(\ell)} \gamma_b  + (-1)^{\ell}\gamma_b \gamma_{a(1)a(2) \cdots a(\ell)} &= \gamma_{a(1)a(2) \cdots a(\ell) b} ,
 \end{align}
where $\check{a(i)}$ represents that the index $a(i)$ is omitted. 
Therefore every multiplication of gamma matrices can be reduced to the linear combination of elements in $\{ \gamma_{a(1) \cdots a(\ell) } \} _{\ell \leq 2n+1}$.
Here $\gamma_{1,2,3,\cdots,2n+1}$ commutes with every generator $\gamma_a$. Its square is 
$\gamma_{1,2,3,\cdots,2n+1}^2 = (-1)^{n(2n+1)} = (-1)^n$. Let us define a matrix $\Gamma$,
\begin{align}
\Gamma &:=  (-1)^{c_n} \gamma_{1,2,3,\cdots,2n+1} ,& c_n &=\left\{ 
\begin{array}{cl} 
0  & n : \mbox{even}\\
1/2  & n : \mbox{odd}
\end{array}  \right. .
\end{align}
The restriction to the eigenspace with respect to $1$ allows us to
treat $\Gamma$ as the identity matrix. 
This restriction implies that all of the $\gamma$s are not linearly independent as follows,
\begin{align}
\gamma_{a(1)a(2) \cdots a(\ell)} \Gamma &=  \frac{1}{(D-\ell)!} (-1)^{\ell(\ell -1)/2 + c_n}  \epsilon_{a(1)a(2) \cdots a(\ell) b(\ell +1) \cdots b(D)} \gamma_{b(\ell +1) \cdots b(D)}  ,  \cr
 \gamma_{a(1)a(2) \cdots a(\ell)} &=  \frac{1}{(D-\ell)!} (-1)^{\ell(\ell -1)/2+c_n}  \epsilon_{a(1)a(2) \cdots a(\ell) b(\ell +1) \cdots b(D)} \gamma_{b(\ell +1) \cdots b(D)}  . 
\end{align}

\subsection{Differential Forms} 
Let us denote the coordinate of the space $M:={\mathbb R}^D$ $x^a$,  $(a=1,2, \cdots , D)$. 

Tangent bundle of $M$ is denoted  $TM$ and cotangent bundle is denoted $T^*M$. 
As an analog of tangent line of plane curve, the fiber of tangent bundle $TM$ is spanned by differentials $\partial / \partial x^a |_x$ where $|_x$ is denoted the valuation of 
derived functions at $x$. 

We can choose $\{ \partial / \partial x^a \}_{a=1}^D$ as generators of sections, globally in a sense of $C(M)$-module. 
Fibers of cotangent bundle $T^*M$ is the dual space of fibers of tangent bundle $TM$. 

Symbol $dx^a$ are denoted their dual generators ruled by the relation $dx^b (  \partial / \partial x^a ) = \delta^b_a $. 
They are generators of the space of sections of cotangent bundle $\Lambda^1:= T^*{\mathbb R}^D$. 
For convention, we use a notation$\Lambda^0 := {\mathbb R} \times M$.
The $s$-th wedge product bundles of $\Lambda^1$ is denoted $\Lambda^s$. 
$dx^{a(1) a(2) \cdots a(\ell)}:=dx^{a(1)} \wedge \cdots \wedge dx^{a(s)}$ are generators of the space of sections of $\Lambda^s$. 
Especially $\Lambda^D$ is rank one vector bundle and the volume form $dv := dx^1 \wedge dx^2 \wedge \cdots \wedge dx^{D} $ generates $\Lambda^D$. 
Hodge dual operator ``$*$" is defined as 
\begin{align}
* dx^{a(1) a(2) \cdots a(\ell )} &:= \frac{1}{(D- \ell ) !} \epsilon_{a(1) a(2) \cdots a(\ell ) b(\ell +1) \cdots b(D)} dx^{b(\ell +1) \cdots b(D)} .
\end{align}
where $\epsilon_{a(1) a(2) \cdots a(\ell ) b(\ell +1) \cdots b(D)}$ is the completely anti-symmetric tensor with values $\epsilon_{123\cdots D}=1$.

Suppose that $\omega$ is a $p$-form. 
\begin{align}
\omega &: = \frac{1}{p!} \sum_{a(1) , a(2) , \cdots , a(p) } \omega_{a(1),a(2), \cdots , a(p)} dx^{ a(1),a(2), \cdots , a(p) } . 
\end{align}
The norm of $p$-form is defined as 
\begin{align}
\omega \wedge * \omega 
&= \frac{1}{p!}  \sum_{ a(1)\cdots a(p) } \omega_{a(1)a(2) \cdots  a(p)}  \omega^{a(1)a(2) \cdots  a(p)}  dv . 
\end{align}
The square of the Hodge dual operator on odd-dimensional spaces is $1$. 
\begin{align}
** \omega 
&= (-1)^{p(D-p)} \omega = \omega . 
\end{align}
The radius of the space $M$ is defined as $r := \sqrt{x^a x^a}$. For non-zero vector $x^a$, the direction vector $e^a$ is defined as 
$e^a := x^a/r$.
Their differentials are written in terms of $x^a$, 
\begin{align}
dr &= \frac{x^a}{r} dx^a , &
de^a 
&= \frac{1}{r} \left(  \delta_{ab} - \frac{x^ax^b}{r^2} \right) dx^b . 
\end{align}
\begin{align}
de^{a(1) a(2) \cdots a(\ell)} 
&= \frac{1}{r^{\ell}} dx^{a(1) a(2) \cdots a(\ell)}  - \sum_{i=1}^{\ell} \frac{x^{a(i)}x^b}{r^{\ell +2}}    dx^{a(1) a(2) \cdots  \stackrel{i}{\check{b}} \cdots a(\ell)} ,\cr
dr \wedge de^{a(1) a(2) \cdots a(\ell)}  
&= \frac{x^b}{r^{\ell +1}} dx^{b a(1) a(2) \cdots a(\ell)} .
\end{align}
Here we would like to compute Hodge duals in terms of $dr$ and $de^a$. In order to simplify the structure of indices, let us consider the contraction with gamma matrices. 
\begin{align}
e &:= e^a \gamma_a  , & e^2 &=1 ,& de &= de^a \gamma_a . 
\end{align}
$de$ is a matrix valued one form. Because $e^2=1$, $de$ and $e$ anticommute with each other, $de e + ede =0 $. 
 The $\ell$-th power of $de$ is denoted $de^{\wedge \ell}$ and 
\begin{align}
de^{\wedge \ell} &= \gamma_{a(1) \cdots a(\ell)} de^{a(1) \cdots a(\ell)}  . 
\end{align}
Let us compute Hodge dual of $ dr \wedge de^{\wedge \ell} $ and $ de^{\wedge \ell} $, 
\begin{align}
 * dr \wedge de^{\wedge \ell} 
 &=  \frac{r^{D-\ell -1} }{r^{\ell }}  \frac{\ell ! (-1)^{(\ell +1)\ell /2+c_n} }{(D-\ell -1)!} e de^{\wedge (D-\ell -1)} ,\cr
  * de^{\wedge \ell} &= 
  \frac{r^{D-\ell -1}}{r^{\ell }}  \frac{\ell !(-1)^{-(D-\ell)(D- \ell -1) /2-c_n} }{(D-\ell -1) !}  
  e dr \wedge de^{\wedge (D- \ell -1)}  .
 \label{eqn:Hodge}
\end{align}

For instance, in the case of $D=5 = 2 \cdot 2 +1 , \ell=3,0$, $c_2=0 $. 
\begin{align}
 * dr \wedge de^{\wedge 3} 
&= \frac{3!}{r^2} ede ,&
 *dr 
&= \frac{r^4}{4!} ede^{\wedge 4} . 
\end{align}

\section{Definition of Models in Odd-dimensional Space}
Let us start talking about $D+1$-dimensional gauge theory. We consider gauge fields belonging to adjoint representation of Spin($D$) and 
scalar fields of vector representation ${\boldsymbol D}$ of Spin($D$). Gauge fields are summarized in a matrix valued one-form and
scalar fields also realized as a matrix valued scalar field by using $\gamma$ matrices, 
$A := A_{\mu}^{ab} dx^{\mu} \gamma_{ab} ,\phi := \phi^a \gamma_a $, $\mu= 0,1,2, \cdots , D$. 
Let us define field strength $F := dA + g A \wedge A $ and covariant derivative of $\phi$, 
$ D \phi := d \phi + g [A , \phi ] $, where $g$ is the gauge coupling constant. 
In general the covariant derivative on the rank $k$ differential form field $\omega$ is defined as
$D \omega = d \omega + g ( A \wedge \omega  - (-1)^k \omega \wedge A )$. 
The covariant derivative of $D \phi$ becomes very simple, $D (D \phi ) = g [ F , \phi ] $. 
 The $k$-the power of the field strength $F$ is denoted 
$F^{\wedge k} := \overbrace{F \wedge F \wedge \cdots \wedge F}^k $. 
In the last part of this section we will discuss some property of a topological current and its corresponding charge.
Until then let us set $A_0=0$ and suppose that all fields are stationary and the operator $*$ is $D$-dimensional spatial Hodge dual operator.  
Because $\gamma_{ab}$ is anti-Hermitian matrix, let us use Hermitian combination
\begin{align}
G &:= \frac{{\bf i} g F}{2\pi}  , & G^{\dag} = G , 
\end{align}
where we adopt a rule which is that the Hermitian conjugation does not change the order of differential forms (for instance $(dx^{ab})^{\dag} = dx^{ab}$).  

In order to explain energies of our model, we use an integer $1 \leq p  \leq n$. $q= n -p$ is also used. 
Let us define the energy density ${\mathcal E}_p$ of our model and energy $E_p$ for stationary configuration of fields. 
\begin{align}
{\mathcal E}_p &:= \| \mu_p G^{\wedge p} \|^2  +   
\|\nu_p \{ D \phi , G^{\wedge q}  \} \|^2  + \lambda V( \phi ) , &
E_p &:= \int {\mathcal E}_p  . 
\end{align}
where $\mu_p$ and $\nu_p$ are positive real parameters and the norm $\| \cdot \|$ of matrix valued differential forms is defined as
$\| X \|^2 := {\rm Tr} X \wedge * X   . $ $V(\phi)$ is the vine bottle potential and we will treat only the case of Prasad-Sommerfeld limit $\lambda \rightarrow 0$. 
$V(\phi)$ stabilizes the scalar field $ \phi^a \phi^a = H_0^2$, where $H_0$ is the vacuum expectation value. From here we will drop the potential because of the PS limit. 
The energy $E_p$ is bound from below by a quantity ${\mathcal Q}_p$ which is defined as follows, 
\begin{align}
E_p & = \int \left\{  \| \mu_p G^{\wedge p} \|^2  +   
\|\nu_p \{ D \phi , G^{\wedge q}  \} \|^2 \right\}  \cr
&= \int  \left\{ \| \mu_p G^{\wedge p} - * \nu_p \varepsilon_p \{ D \phi , G^{\wedge q}  \} \|^2 + 2\varepsilon_p  \mu_p \nu_p {\rm Tr} G^{\wedge p} \wedge  \{ D \phi , G^{\wedge q}  \} \right\} \cr
& \geq 2\varepsilon_p  \mu_p \nu_p \int_{S^{2n}_{\rm phys}}   {\rm Tr} \phi G^{\wedge n}  =: {\mathcal Q}_p .
\end{align}
Here $\varepsilon_p$ is a sign $\pm$ and  the $2n$-dimensional sphere $S^{2n}_{\rm phys}$ in the last line is, so-called, ``spatial boundary" 
which is a sphere with very large radius $R$, whose center is sitting near the center of mass of the system. 
$\{ A , B \}:= AB + BA$ stands for the anti commutation relation 
The inequality attains the equality if the following equation is satisfied, 
\begin{align}
\mu_p G^{\wedge p} &= \nu_p \varepsilon_p * \{ D \phi , G^{\wedge q}  \} . 
\end{align}
This is the Bogomol'nyi equation of this model. 

In order to confirm that the quantity ${\mathcal Q}_p = 2\varepsilon_p  \mu_p \nu_p \int_{S^{2n}_{\rm phys}}   {\rm Tr} \phi G^{\wedge n} $ is topological, 
let us consider the following topological charge which was introduced by Arafune, Freund and Goebel \cite{Arafune:1974uy}.  
Tchrakian and Zimmerschied discuss about their generalization \cite{Tchrakian:1978sf}.  
Let us recover the time direction and time dependence. 
\begin{align}
{\mathcal F}_n &:=   {\rm Tr}  \hat{\phi}  f_n( F , D \hat{\phi} )  = {\rm Tr}  \hat{\phi}  \left\{ \frac{1}{(4g)^{n}}  D\hat{\phi}^{2n} +{\mathcal R}_n(D \hat{\phi}  , F) \right\}  , & 
\hat{\phi} &:= \frac{\phi}{ |\phi | },& | \phi | &=\sqrt{ \phi^a \phi^a } ,  
\end{align}
Here $f_n$ consists of $2n$-forms. Of course these quantities are well defined only in the region where $ | \phi | \neq 0$. 
At least for configurations with finite energy, at points $x$which is far from such a configuration $\phi(x) \sim H_0$. Otherwise the energy should diverge. 
Therefore we will work around such an asymptotic region. 
The remainder term ${\mathcal R}_n$ is determined as the equality 
\begin{align}
d {\mathcal F}_n = {\rm Tr} (D \hat{\phi})^{2n+1}/(4g)^{n} .
\label{eqn:Fn}
\end{align} 
For instance, ${\mathcal F}_2$ and ${\mathcal F}_3$ are
\begin{align}
 {\mathcal F}_2 &= {\rm Tr} \hat{\phi} \left( F + \frac{1}{4g} D \hat{\phi} \wedge D \hat{\phi}   \right)^2   , \cr
 {\mathcal F}_3  &=  {\rm Tr} \hat{\phi}  \left( F^3 + \frac{1}{8g} ( D\hat{\phi}^2 F^2 + F D\hat{\phi}^2 F + F^2 D\hat{\phi}^2 
+  D \hat{\phi} F D \hat{\phi}  F + D \hat{\phi} F^2 D \hat{\phi} +F D \hat{\phi} F D \hat{\phi}   )  \right. \cr
& \left.+ \frac{3}{(4g)^2} D\hat{\phi} ^4 F 
+ \frac{1}{(4g)^3}  D\hat{\phi}^6  \right) . 
 \end{align}
As is shown, ${\mathcal F}_3$ does not factorize.  
Can we find the solution ${\mathcal F}_n$ for the equation (\ref{eqn:Fn}) in general $n$? Yes, we can. 
The derivative $d$ can be replace as covariant derivative $D$ inside trace:  $d  {\rm Tr}  X = {\rm Tr} D X $. 
The covariant derivative of $F$ vanishes (Bianchi identity). Therefore descents appear as terms including $D^2 \hat{\phi}$ which can be replaced by $g [ F , \hat{\phi} ]  $. 
It is nontrivial that we can erase all terms including several $\hat{\phi}$. 
\begin{align}
d {\rm Tr}\hat{\phi}  (D \hat{\phi})^{2n}/(4g)^{n} &= \frac{1}{(4g)^{n}}  {\rm Tr} \left\{  (D \hat{\phi}   )^{2n+1} +   \hat{\phi} D (D \hat{\phi}   )^{2n} \right\}\cr
 \frac{1}{(4g)^{n}}  {\rm Tr} \left\{  \hat{\phi} D (D \hat{\phi}   )^{2n} \right\} &=  
\frac{1}{(4g)^{n}}  {\rm Tr} \left\{  \hat{\phi} \sum_{l=0}^{2n-1}  (-1)^l (D \hat{\phi}   )^{l} (D^2 \hat{\phi})  (D \hat{\phi}   )^{2n-l-1} \right\} \cr
&= - \frac{n}{(4g)^{n-1}}  {\rm Tr}   (D \hat{\phi}   )^{2n-1} F  . 
\label{eq:remainder1}
\end{align}
Therefore let us suppose that ${\mathcal R}_n(D \hat{\phi}  , F) $ starts from the term 
\begin{align}
{\mathcal R}_n^{(1)}(D \hat{\phi}  , F) &:= {\mathcal C}_1 (D \hat{\phi}   )^{2n-2} F , & {\mathcal C}_1 &:= \frac{n}{(4g)^{n-1}}  . 
\end{align}
 which yields a counter term of the descent 
$ \frac{1}{(4g)^{n}}  {\rm Tr} \left\{  \hat{\phi} D (D \hat{\phi}   )^{2n} \right\} $. Let us continue the computation. $d {\rm Tr} \hat{\phi} {\mathcal R}_n^{(1)}(D \hat{\phi}  , F)$ becomes
\begin{align}
 {\mathcal C}_1 d  {\rm Tr}  \hat{\phi}   (D \hat{\phi}   )^{2n-2 } F &=   \frac{n}{(4g)^{n-1}}  {\rm Tr} D \left\{   \hat{\phi}   (D \hat{\phi}   )^{2n-2 } F  \right\} \cr
  &= {\mathcal C}_1  {\rm Tr}  \left\{     (D \hat{\phi}   )^{2n-1} F +  \hat{\phi}  D(D \hat{\phi}   )^{2n-2}  F \right\} . 
\end{align}
The first term is the counter term of (\ref{eq:remainder1}) and now we have interest in the remainder,  
\begin{align}
{\mathcal C}_1  {\rm Tr}  \left\{  \hat{\phi}  D(D \hat{\phi}   )^{2n-2}  F \right\}  &= {\mathcal C}_1 {\rm Tr}  \left\{  \hat{\phi} 
\sum_{l=0}^{2n-3} (-1)^l  (D \hat{\phi}   )^{l}   D^2 \hat{\phi}   (D \hat{\phi}   )^{2n-3-l}  F \right\} \cr
&= g {\mathcal C}_1  {\rm Tr}  \left\{  \hat{\phi} 
\sum_{l=0}^{2n-3} (-1)^l  (D \hat{\phi}   )^{l}   (F  \hat{\phi} - \hat{\phi}  F)   (D \hat{\phi}   )^{2n-3-l}  F \right\} .
\end{align}
Let us split the commutator part and move two $\hat{\phi}$s,  
\begin{align}
&= g {\mathcal C}_1  {\rm Tr}  \left\{ 
\sum_{l=0}^{2n-3}  (D \hat{\phi}   )^{l}  \hat{\phi}   F  \hat{\phi}   (D \hat{\phi}   )^{2n-3-l}  F
- \sum_{l=0}^{2n-3}  (D \hat{\phi}   )^{l}     F   (D \hat{\phi}   )^{2n-3-l}  F \right\}  .
\label{eq:remainder2}
\end{align}
The second term in (\ref{eq:remainder2}) can be derived from some combination ${\rm Tr} \hat{\phi} Y(F, D \hat{\phi} )$, where $Y$ is some polynomial.  
Let us check that the first term vanishes. 
\begin{align}
& g {\mathcal C}_1 {\rm Tr}  \left\{ 
\sum_{l=0}^{2n-3}  (D \hat{\phi}   )^{l}  \hat{\phi}   F  \hat{\phi}   (D \hat{\phi}   )^{2n-3-l}  F \right\} \cr
 &= \frac{g {\mathcal C}_1}{2}   {\rm Tr}  \left\{ 
\sum_{l=0}^{2n-3}  (D \hat{\phi}   )^{l}  \hat{\phi}   F  \hat{\phi}   (D \hat{\phi}   )^{2n-3-l}  F  -
 \sum_{l=0}^{2n-3}   (D \hat{\phi}   )^{l} \hat{\phi}  F \hat{\phi}(D \hat{\phi}   )^{2n-3-l}    F     \right\} =0 . 
\end{align}
The following term is the descent term from ${\mathcal R}_n^{(1)}(D \hat{\phi}  , F)$, 
\begin{align}
 g {\mathcal C}_1 {\rm Tr}  \left\{ \sum_{l=0}^{2n-3}  (D \hat{\phi}   )^{l}     F   (D \hat{\phi}   )^{2n-3-l}  F \right\} . 
\label{eqn:r2d}
\end{align}
From this result, let us define next term of ${\mathcal R}_n(D \hat{\phi}  , F)$  as follows, 
\begin{align}
{\mathcal R}_n^{(2)}(D \hat{\phi}  , F) &:=
{\mathcal C}_2 \sum_{k+l+m=2n-4}   (D \hat{\phi}   )^{k}     F   (D \hat{\phi}   )^{l }  F (D \hat{\phi}   )^{m}   ,
\end{align}
Here ${\mathcal C}_2$ is determined by comparing the coefficient of ${\rm Tr}(D \hat{\phi}   )^{2n-3}  F^2$ in (\ref{eqn:r2d}) and 
that in $ d {\rm Tr} \hat{\phi}  {\mathcal R}_n^{(2)}(D \hat{\phi}  , F) $. 
\begin{align}
d {\rm Tr} \hat{\phi}  {\mathcal R}_n^{(2)}(D \hat{\phi}  , F)  
&= {\mathcal C}_2 {\rm Tr} \sum_{k+l+m=2n-4}    (D \hat{\phi}   )^{k+1}     F   (D \hat{\phi}   )^{l }  F   (D \hat{\phi}   )^{m}  .  \cr
& +{\mathcal C}_2 {\rm Tr} \sum_{k+l+m=2n-4}  \hat{\phi} D \left\{  (D \hat{\phi}   )^{k}     F   (D \hat{\phi}   )^{l }  F   (D \hat{\phi}   )^{m}   \right\} . 
\label{eqn:dr2} 
\end{align}
 (\ref{eqn:r2d})  includes  $2 g {\mathcal C}_1 {\rm Tr}  (D \hat{\phi}   )^{2n-3}     F^2$
and from (\ref{eqn:dr2}) the following term is read off, 
\begin{align}
& {\mathcal C}_2 {\rm Tr} \left(  \sum_{k+m=2n-4}    (D \hat{\phi}   )^{k+1}     F^2   (D \hat{\phi}   )^{m}  \right) 
=(2n-3)  {\mathcal C}_2 {\rm Tr}    (D \hat{\phi}   )^{2n-3}     F^2   .
\end{align}
Therefore we obtain the coefficient  ${\mathcal C}_2= {n}/{2(2n-3)(4g)^{n-2}} $. 
The remaining term in (\ref{eqn:dr2})  reduce to the form like $Y(D \hat{\phi} , F ) $. 
\begin{align}
&{\mathcal C}_2 {\rm Tr} \sum \hat{\phi} D \left\{  (D \hat{\phi}   )^{k}     F   (D \hat{\phi}   )^{l }  F   (D \hat{\phi}   )^{m}   \right\}  \cr
&=  - 2 g {\mathcal C}_2 {\rm Tr} \sum_{k+l+m+u=2n-5} 
  (D \hat{\phi}   )^{k}         F   (D \hat{\phi}   )^{l} 
   F   (D \hat{\phi}   )^{m }  F   (D \hat{\phi}   )^{u} .
\end{align}
The remaining terms which include $\hat{\phi}$ without derivative vanish. 
These result suggest  the following general definition:
\begin{align}
 {\mathcal R}_n^{(s)}(D \hat{\phi}  , F) 
&:= {\mathcal C}_s  \sum_{k(1), \cdots, k(s+1), \atop \sum k(i)=2(n-s)}  
 (D \hat{\phi}   )^{k(1)} F  (D \hat{\phi}   )^{k(2)} F (D \hat{\phi}   )^{k(3)} \cdots   (D \hat{\phi}   )^{k(s)} F  (D \hat{\phi}   )^{k(s+1)}  . 
\end{align}
Let us follow the same procedure as before. The exterior derivative of ${\rm Tr} \hat{\phi} {\mathcal R}_n^{(s)}(D \hat{\phi}  , F)$ is 
\begin{align}
d {\rm Tr} \hat{\phi} {\mathcal R}_n^{(s)}(D \hat{\phi}  , F)  &= {\rm Tr} \left\{ D \hat{\phi} {\mathcal R}_n^{(s)}(D \hat{\phi}  , F)   + \hat{\phi} D {\mathcal R}_n^{(s)}(D \hat{\phi}  , F) \right\}  .
\end{align}
$ {\rm Tr} (D \hat{\phi}) {\mathcal R}_n^{(s)}(D \hat{\phi}  , F)  $ is the counter term of the descents from $  {\rm Tr}  \hat{\phi} D {\mathcal R}_n^{(s-1)}(D \hat{\phi}  , F)  $. 
\begin{align}
&  {\rm Tr}  \hat{\phi} D {\mathcal R}_n^{(s)}(D \hat{\phi}  , F) \cr
   &= -  2 g  {\mathcal C}_s  {\rm Tr}  \sum_{l(1), \cdots, l(s+2), \atop \sum l(i)=2(n-s)-1}    
   (D \hat{\phi}   )^{l(1)} F  (D \hat{\phi}   )^{l(2)} \cdots F  (D \hat{\phi}   )^{l(s+1)} F  (D \hat{\phi}   )^{l(s+2)}  
\end{align}
Therefore we obtain that the following sum gives correct ${\mathcal F}_n$ which satisfies $d {\mathcal F}_n = {\rm Tr} (D \hat{\phi} )^{2n+1}/(4g)^n $. 
\begin{align}
{\mathcal R}_n(D \hat{\phi}  , F)  &:= \sum_{s=1}^{n}    {\mathcal R}_n^{(s)}(D \hat{\phi}  , F)   . 
\end{align}
Let us define the coefficients  $ {\mathcal C}_s $. By comparing coefficients of $(D \hat{\phi} )^{2(n-s)+1}F^s$ we obtain 
\begin{align}
 {\mathcal C}_{s} &:=   
\frac{ (2n-2s-1)!!n!  }{ (2n-s)!  (2g)^{n-s} } ,& &(2 \leq s \leq n ) .
\end{align}
This result corresponds with ${\mathcal F}_2$ and  ${\mathcal F}_3$ shown previously. 
\begin{align}
n&=2 ,& {\mathcal C}_{2}&=   \frac{ (4-4-1)!!2!  }{ (4-2)!  }= 1, && \cr
n&= 3, &  {\mathcal C}_{3} &= \frac{ (6-6-1)!!3!  }{ (6-3)!  (2g)^{3-3} } = 1 , & {\mathcal C}_{2} &= \frac{ (6-4-1)!!3!  }{ (6-2)!  (2g)^{3-2} } =  \frac{1}{8g} . 
\end{align}
In general, $ {\mathcal C}_{n} = 1$ and this means that  ${\mathcal R}_n^{(n)}(D \hat{\phi}  , F) = F^n$. 

The covariant derivative in the term ${\rm Tr} (D \hat{\phi})^{2n+1}/(4g)^{n}$ can be replaced by just $d$. 
 $\hat{\phi}$ and $d \hat{\phi}$ anticommute with each other.  Similarly, $\hat{\phi}$ and $[A ,  \hat{\phi}] $ anticommute with each other. 
 Therefore the commutator terms vanish
\begin{align}
{\rm Tr}[A , \hat{\phi} ]  (D \hat{\phi})^{2k} (d \hat{\phi})^{2l} &= {\rm Tr}  \left(  A  \hat{\phi}   (D \hat{\phi})^{2k} (d \hat{\phi})^{2l}   -  \hat{\phi}  A   (D \hat{\phi})^{2k} (d \hat{\phi})^{2l}   \right) \cr
&= {\rm Tr}  \left(   \hat{\phi} A    (D \hat{\phi})^{2k} (d \hat{\phi})^{2l} -  \hat{\phi}  A  (D \hat{\phi})^{2k} (d \hat{\phi})^{2l}   \right)    =0 , 
\end{align}
Therefore ${\rm Tr} (D \hat{\phi})^{2n+1} =  {\rm Tr} (d \hat{\phi})^{2n+1} $. 
This ensures that the differential form ${\mathcal F}_n$ can be separated into two parts, 
\begin{align}
{\mathcal F}_n &= d {\mathcal H}_n +  \frac{1}{(4g)^{n}} {\rm Tr}  \hat{\phi}    d\hat{\phi}^{2n}  . 
\end{align}
For example,  ${\mathcal F}_2$ is written as follows,
\begin{align}
 {\mathcal F}_2 &=  d {\mathcal H}_2 + \frac{1}{16g^2} {\rm Tr} \hat{\phi}   d \hat{\phi}^{\wedge 4} \cr
{\mathcal H}_2 &: =   \frac{1}{2} {\rm Tr} \left\{    \frac{1}{g} {\rm Tr} \hat{\phi} d \hat{\phi}^2 A    
+  \left(  \hat{\phi}  (  dA A   + A dA )  +  \hat{\phi} d\hat{\phi} A\hat{\phi} A     \right)
 + g \left(  \hat{\phi}   A^3 +\frac{1}{3} \hat{\phi}A\hat{\phi} A\hat{\phi}A   \right) \right\} .
\end{align}
Let us define the topological charge 
\begin{align}
Q_{\rm top}^{(2n)} &:=\frac{1}{{\rm vol}  (S^{2n}|_{r=1} ) }  \int_{S^{2n}_{\rm phys}} \hat{\phi} ( d \hat{\phi} )^{\wedge 2n} . 
\end{align}
This charge counts Brower degree of the map $x \mapsto \hat{\phi}(x)$. 
Suppose that configuration concentrate around the origin of the space ${\mathbb R}^{2n+1}$. 
A ball $B( R_0) := \{  x \in {\mathbb R}^{2n+1} | |x|\leq R_0 \} $ is inside of the sphere whose center is the origin and radius is $R_0$. 
Then we obtain the following equation. 
\begin{align}
\int_{ B( R_0) }  d {\mathcal F}_n &= \int_{S^{2n}|_{r=R_0} }  \frac{1}{(4g)^{n}} {\rm Tr}  \hat{\phi}    d\hat{\phi}^{2n}  =\frac{{\rm vol}  (S^{2n}|_{r=1} ) }{(4g)^{n}}  Q_{\rm top}^{(2n)} . 
\end{align}
We only consider configurations with finite energy. 
In asymptotic region,  $D\phi$ should  vanishes. Therefore we can estimate $ {\mathcal F}_n \sim   {\rm Tr} \hat{\phi} F^n \sim {\rm Tr} 
\phi F^n /H_0 $. The last expression proportional to the charge ${\mathcal Q}_p$. Therefore we conclude that ${\mathcal Q}_p$ is topological.

\section{Hedge-Hog ansatz and Abel's differential Equation}
We would like to consider Hedge-Hog solutions of the Bogomol'nyi equation of our model. 
Convergence of the energy integral requires that $F = o( r^{-D/2p})$ and $D \phi = o(r^{-D(1-q/p)/2})$.  
This shows that in the case $q > p$, the energy may converges even though $D \phi$ diverges. 
However for such a solution the contribution from the vine bottle potential cannot be finite and the case should be excluded. 
Now we want to consider the case where $\phi \rightarrow H_0$ in the asymptotic region $r >> 0$.  Therefore we require $D \phi \rightarrow 0$ in any case. 
Let us make an ansatz,
\begin{align}
\phi &:= H_0 U(r) e , &
A &:= \frac{1-K(r)}{2g} ede , 
\end{align}
where $U(r)$ and $K(r)$ are functions of $r$. $e$ and $de$ anticommute with each other, 
$ede +dee=0$. Boundary conditions are given by $U(0)=0,K(0)=1,U(\infty)=1$ and $K(\infty)=0$. 
The covariant derivative of $\phi$  and field strength $F$ are written as
\begin{align}
D \phi &= H_0 ( KU de + U' edr) ,&
F &= \frac{1-K^2}{4g} de \wedge de  - \frac{K'}{2g} edr \wedge de .
\end{align}

The powers of field strength are 
\begin{align}
F^{\wedge p} &= \left( \frac{1-K^2}{4g} \right)^p de^{\wedge 2p} - p \frac{K'}{2g} \left( \frac{1-K^2}{4g} \right)^{p-1} edr \wedge de^{\wedge 2p-1}  . 
\end{align}
There are two cases $p=n, q=0$, and $q \neq 0$. It is better to discuss separately. Through this article, we do not consider the case $p=0$. 
\subsection{ $p=n,q=0$}
In this case $\{ D \phi , G^0 \} = \{ D \phi , 1 \} = 2 D \phi$. Let us use simplified notation $\mu:= \mu_n, \nu := 2 \nu_n$. 
\begin{align}
{\mathcal E}_n &:= \| \mu G^{\wedge n} \|^2  +   
\|\nu D \phi \|^2  + \lambda V( \phi ) , &
E_n &:= \int {\mathcal E}_n  . 
\end{align}
The corresponding Bogomol'nyi equation is
\begin{align}
\mu G^{\wedge n} &= \nu \varepsilon_n *  D \phi . 
\end{align}
%
%
For the Hedge-Hog ansatz it reduces to a system of ordinary differential equations. 
\begin{align}  
\mu \left( \frac{{\bf i}g}{2\pi} \right)^n  \left( \frac{1-K^2}{4g} \right)^{n-1} \left( \frac{1-K^2}{4g} de^{\wedge 2n} 
- n \frac{K'}{2g}  edr \wedge de^{\wedge 2n-1}   \right)
&= \nu \varepsilon_n * H_0 ( KU de + U' edr ) .
\label{eqn:qe0:bps1}
\end{align}
Remember the equation (\ref{eqn:Hodge}) and 
\begin{align}
* edr &= r^{2n} \frac{ (-1)^{c_n} }{(2n)!} de^{\wedge 2n} ,&
* de  &= \frac{r^{2n-2}}{(2n-1)!} (-1)^{n -c_n} e dr \wedge de^{2n-1} . 
\end{align}

The equations (\ref{eqn:qe0:bps1}) are equivalent to
\begin{align}
\mu \left( \frac{{\bf i}g}{2\pi} \right)^n  \left( \frac{1-K^2}{4g} \right)^{n}   &= \nu \varepsilon_n  H_0 r^{2n} \frac{ (-1)^{c_n} }{(2n)!}     U' ,\cr
- \mu n \left( \frac{{\bf i}g}{2\pi} \right)^n  \left( \frac{1-K^2}{4g} \right)^{n-1}  \frac{K'}{2g} &= \nu \varepsilon_n  H_0 \frac{r^{2n-2}}{(2n-1)!} (-1)^{n-c_n}  KU . 
\end{align}
Here we would like to consider the situation where the functions $U$ and $K$ monotonically change. 
By using the expression $n=2([n/2]+c_n)$, the sign is determined as $\varepsilon_n = (-1)^{[n/2]} $ and $n-2 c_n \equiv 0 (\mbox{mod} 2)$. 
The Bogomol'nyi equation becomes
\begin{align}
\alpha ( 1-K^2 )^{n}   &=  r^{2n}      U' ,& - \alpha   ( 1-K^2 )^{n-1}  K' &=  r^{2n-2}   KU ,& 
\alpha &= \frac{\mu (2n)!}{\nu   H_0 (8 \pi)^n}  . 
\end{align}
By using a new function $f=K^2$ we obtain
\begin{align}
\alpha ( 1-f )^{n}   &=  r^{2n}      U' ,& - \frac{\alpha}{2}   ( 1-f )^{n-1}  f' &= (-1)^n r^{2n-2}   f U .  
\end{align}
Suppose that $1-f$ and $U$ damp  as fast as powers of $r$. 
\begin{align}
1-f &= O(r^{\lambda_1}) , & 
U &= O(r^{\lambda_2}) ,& \lambda_1 ,\lambda_2 > 0 . 
\end{align} 
\begin{align}
n \lambda_1 - \lambda_2 &= 2n-1 , & 
\lambda_2 &= n \lambda_1 - (2n-1) , & 
2-  \frac{1}{n} &\leq \lambda_1 . 
\end{align}
This shows that $f' \rightarrow 0$. 
 The boundary conditions for $f$ at $r=0$ are $f(0)=1, f'(0)=0$. 

Around $r=\infty$, $f$ goes to $0$ and $U$ goes to $1$. 
\begin{align}
 U' & \sim \frac{\alpha}{r^{2n}}  ,& U & \sim  - \frac{\alpha}{2}   \frac{ f'}{r^{2n-2} f}  .  
\end{align}
 $U$ should converge to $1$. Therefore $ (\ln f)' = - (2/\alpha) r^{2n-2} (1+ O(1/r)) $. 
 The asymptotic expansions of $U$ and $f$ are 
 \begin{align}
 U &=   1-  \frac{\alpha}{(2n-1)r^{2n-1}}  + O(r^{-2n}) , & f &= C \exp \left( - \frac{2}{\alpha (2n-1)} r^{2n-1} (1+ O(1/r)) \right)   .
 \end{align}
 In this case $f(\infty)=0,  f'(\infty)= 0 $. 
By erasing the function $U$,  we obtain
\begin{align}
2f ( 1-f )   &= 2(n-1)   r f ' 
- r^2 f ''  +  \frac{1+ (n-2)f } {f(1-f)} r^2(f ')^2  . 
\end{align}

Let us define new variable $\tau = \ln (r/R)$, where $R:= \alpha^{1/(2n-1)} $. $\dot{f}$ represents $\tau$-derivative of $f$. 
Then we obtain an autonomous equation, 
\begin{align}
  \ddot{f} &=  \frac{1+(n-2)f}{f(1-f)}  (\dot{f})^2 +   (2n-1)  \dot{f}   -  2f(1-f)   . 
\end{align}
Because the equation is autonomous, the degree of the differential equation can be decreased by the substitution, $x:=f, ~y\equiv y(x) :=\dot{f}$, 
\begin{align}
 y\frac{dy}{dx}&=  \frac{1+(n-2)x}{x(1-x)}  y^2 +   (2n-1) y   -  2x(1-x)   . 
\end{align}
This is Abel's differential equation. 
Fixed points are 
\begin{align}
(x,y) &= ( 0, 0) , ( 1 , 0)  . 
\end{align}
Boundary conditions for $\tau= - \infty$ are $(1,0)$ and those for $\tau=+ \infty$ are $(0,0)$. 

\subsection{$q \neq 0$} 
Suppose that $q \neq 0$. The Bogomol'nyi equation is 
\begin{align}
\mu_p G^{\wedge p} &= \nu_p \varepsilon_p * \{ D \phi , G^{\wedge q}  \} . 
\end{align}
The anticommutator becomes
\begin{align}
\{ D \phi , F^{\wedge q} \} 
&= 2 H_0 \left( \frac{1-K^2}{4g} \right)^{q-1} \left\{ \frac{1-K^2}{4g} KU de^{\wedge 2q+1}  \right.\cr
& \left. +\left( \frac{1-K^2}{4g} \right)^{-p+1} \left( \left( \frac{1-K^2}{4g} \right)^p  U  \right)' edr  \wedge de^{\wedge 2q} \right\}  . 
\end{align}
Again, let us use the function $f:=K^2$. 
From the Bogomol'nyi equation, 
we obtain 
\begin{align}
 \beta_p   ( 1-f )^{2p-q} &=  \varepsilon_p   \frac{ (-1)^{n+c_n + n/2} }{r^{2(q-p) }}  \left( (1-f)^p  U  \right)'  \cr
- \beta_p  p f' (1-f)^{p-q-1}  &=   \varepsilon_p   \frac{(-1)^{ -c_n+ n/2} (2q+1)2p }{ r^{2(q-p)+2} }  fU ,
\end{align}
where the coefficient can be gathered up together as 
\begin{align}
\beta_p &:= \frac{ \mu_p (2p)!} {2 H_0\nu_p  (2q)!( 8\pi )^{p-q}} .
\end{align}
The sign is determined as $ \varepsilon_p :=  (-1)^{n+c_n + n/2}$. Then the differential equations become
\begin{align}
 \beta_p  ( 1-f )^{2p-q} &=      \frac{ 1}{r^{2(q-p) }} \left( (1-f)^p  U  \right)' ,&
- \beta_p   p f' (1-f)^{p-q-1}  &=    \frac{(2q+1)2p }{  r^{2(q-p)+2}   }   fU  .
\end{align}
Suppose that $1-f$ and $U$ damp  as fast as powers of $r$. 
\begin{align}
1-f &= O(r^{\lambda_1}) , & 
U &= O(r^{\lambda_2}) ,& 
\lambda_1 , \lambda_2 &> 0 . 
\end{align} 
We obtain conditions for these parameters $\lambda_1,\lambda_2$, 
\begin{align}
\lambda_1 (p-q)  - \lambda_2 &= 2(p-q)   -1 , &
\lambda_2 &= ( \lambda_1 -2 )(p-q) +1 ,& 
2- \frac{1}{p-q} & \leq \lambda_1 , 
\end{align}

Around $r=\infty$, $f \rightarrow 0$ and $U \rightarrow 1$. 
\begin{align}
 U' & \sim  \frac{\beta_p}{ r^{2(p-q)} }  ,&
   (\ln f)'  &\sim   -   \frac{2(2q+1)}{\beta_p}   r^{2(p-q-1)}     .
\end{align}
Therefore the condition $p > q$ is required for $| \phi | \rightarrow H_0$ under the limit $r \rightarrow +\infty$.  
Expansions around $r = \infty$ are
\begin{align}
U & = 1-  \frac{\beta_p}{\{ 2(p-q)-1\}  r^{2(p-q)-1}} + O(r^{-2(p-q)}) ,\cr
f &= C \exp \left\{ - \frac{ 2(2q+1) }{\beta_p  \{2(p-q)-1\}}  r^{2(p-q)-1} (1+ O(1/r))\right\}   .
\label{eq:behavioraroundinfinity}
\end{align}
The boundary conditions at $r=\infty$ are $f(\infty)=0, f'(\infty)=0$. The typical scale is given by $R_p := \beta^{1/\{2(p-q)-1\}} $. New variable $\tau$ is defined as $\tau = \ln (r/R_p)$. 
In the same manner as the previous case,  we obtain 
\begin{align}
  \ddot{f}  &=   \frac{1 +(2p-q-2) f }{f(1-f)} (\dot{f})^2   +    \{ 2 (p-q) -1 \}   \dot{f}    -    2(2q+1) f   ( 1-f )  . 
\end{align}
where $\dot{f}$ represents $\tau$-derivative of $f$. 
Coordinates  $x:=f, y \equiv y(x) := \dot{f}$ lead us to 
\begin{align}
 y \frac{dy}{dx}  &=   \frac{1 +(2p-q-2) x }{x(1-x)}y^2   +    \{ 2 (p-q) -1 \}   y   -    2(2q+1) x   ( 1-x )  . 
\end{align}
The equation is Abel's differential equation, too.

\section{Spaces of Jets}
In the previous section, we obtain two series of Abel's differential equations. 
\begin{align}
F_p(x,y, v) &:= x(1-x) y v -  ( 1 + A_p x ) y^2 - B_p  x(1-x)  y  + C_p x^2 (1-x)^2 , \cr
v &:= \frac{dy}{dx} ,~~
F_p(x,y, v) =0 ,
\end{align}
where parameters $A_p,B_p,C_p$ are 
\begin{align}
A_n &:= n-2 , & B_n &:= 2n-1 , & C_n &:= 2 , \cr
A_p &:= 3p-n-2, & B_p &:= 4p-2n-1 ,& C_p &:= 4n-4p+2 ,~~(p \leq n-1) . 
\end{align}
The three-dimensional space with coordinates $(x,y,v)$ is called the space of 1-jets of the function $y(x)$ \cite{Arnold:1983}. 
Let us consider the surface defined by the equation $F_p(x,y,v)=0$. 
Singularities on this surface is given as a algebraic set given by $\nabla F_p = ( \partial_x, \partial_y, \partial_v ) F_p = 0$ and $F_p=0$. 
 The equation $dF_p/dv=0$ shows that there are three cases. 
\begin{enumerate}
\item $x=0$ ~~ In this case, ${\partial F_p}/{\partial y} =  - 2 y =0$ shows that the equation ${\partial F_p}/{\partial x}=0$ and the defining equation $F_p=0$ are automatically satisfied. Therefore This singularity is a line $x=y=0$.  
\item $x=1$ ~~ We obtain $y=0$ and the equation $\partial F_p / \partial x=  0$ is satisfied. This singularity is a line $x=1, y=0$. 
\item $y=0$ ~~ In order to find new singularity, suppose that $x \neq 0,1$. $v= B_p$ and $x=1/2$. 
\end{enumerate}
%
We would like to remove the singular lines $x=y=0$ and $x=1,y=0$. 
 Around the line $(x,y,p)=(0,0,p)$ the function $F_p$ becomes
\begin{align}
F_p(x,y, v) & \sim x y v -   y^2 - B_p  x  y  + C_p x^2 . 
\end{align}
This suggest us that if $y$ is proportional to $x$ it can be divided by $x^2$. 
Let us use new coordinate $\tilde{y}$ defined as $y=: x\tilde{y}$. The definition equation is rewritten as
\begin{align}
 F_p(x,y,v) &= x^2 \left\{ (1-x) \tilde{y} v -  ( 1 + A_p x ) \tilde{y}^2 - B_p (1-x)  \tilde{y}  + C_p (1-x)^2 \right\} .  
\end{align}
This transformation is called blow-up. Let us drop the over-all term $x^2$. 
We obtain new surface defined by
\begin{align}
\tilde{F}_p(x,\tilde{y}, v)  &:= (1-x) \tilde{y} v -  ( 1 + A_p x ) \tilde{y}^2 - B_p (1-x)  \tilde{y}  + C_p (1-x)^2 . 
\end{align}
This equation still have a singular line $x=1,\tilde{y}=0$. 
Suppose that a point $(x,y)$ is close to $(1,0)$.  
Let us re-parametrize as $x=1+u$ and suppose that these parameters are small $u \sim 0 , \tilde{y} \sim 0$.
 With these coordinates the polynomial $\tilde{F}_p(x,\tilde{y}, v) $ is 
\begin{align}
\tilde{F}_p(x,\tilde{y}, v)  = u \tilde{y} v -  ( 1 + A_p  ) \tilde{y}^2 - B_p u  \tilde{y}  + C_p u^2 . 
\end{align}
Therefore let us consider a transformation $\tilde{y}= (1-x)  w$. After that, $\tilde{F}_p(x,\tilde{y}, v) $ factorizes and we obtain new surface defined by $H_p(x,w,v) $. 
\begin{align}
\tilde{F}_p(x,\tilde{y}, v)  &= (1-x)^2 \left\{  w v -  ( 1 + A_p x )  w^2 - B_p   w  + C_p \right\} ,\cr
H_p(x,w,v) &:= w v -  ( 1 + A_p x )  w^2 - B_p   w  + C_p  . 
\end{align}
The surface defined by $H_p$ is smooth. 
However this surface does not relate to differential equation directly. 
Let us apply the transformation $y= x(1-x)w$ to the original differential equation. 
%
\begin{align}
F_p\left( x,y,\frac{dy}{dx} \right) &= x^2(1-x)^2 \left\{ x(1-x)w \frac{dw}{dx} + (1-2x)w^2 - (1+A_p x)w^2 - B_p w  +C_p  \right\}  .
\end{align}
We have no interest in the solution $x \equiv 0,1$, therefore we will consider the equation, 
\begin{align}
G_p(x,w,z) &:= x(1-x)w z  - (2+A_p )x w^2 - B_p w  +C_p  ,& G_p\left( x,w, \frac{dw}{dx} \right)&=0 . 
\end{align}
%
Here $A_p,B_p,C_p$ are integer and $B_p$ is odd. Therefore this surface is regular as an affine curve. 

Before discussing about this equation, we will study some property of this surface.  
Let us consider embedding into the complex projective space ${\mathbb C}P^3$. 
Homogeneous coordinates $[X:Y:Z:W]$ of ${\mathbb C}P^3$ are related to the original coordinates by
$x := {X}/{W}, 
w :={Y}/{W} , 
z :={Z}/{W} $. 
By substitution into $G_p(x,w,z)$ and multiplying $W^4$, we obtain a homogeneous polynomial, 
\begin{align}
K_p(X,Y,Z,W)&:= X(W-X) Y Z  - (2+A_p )X Y^2W - B_p YW^3   +C_pW^4  . 
\end{align}
This polynomial is irreducible and this polynomial defines a variety. 
In addition, the degree of  this polynomial is four. 
However this projective surface has singularities. There are a line singularity $[0:Y:Z:0]$ and a point singularity $[X:0:0:0]$. 
Both are sitting on the four-cycle ${\mathbb C}P^2$ defined by $W=0$. Therefore this surface is not a K3 surface.  

\section{Condition for the label $p$ and Asymptotic Solution}
\begin{figure}[tb]
 \psfrag{sintheta}{\huge $\sin 3 \theta$}  
 \psfrag{n}{\huge $n$}
 \psfrag{xi0}{\huge $\xi_0$}
 \begin{minipage}{80mm}
\rotatebox{270}{\resizebox{55mm}{!}{ \includegraphics{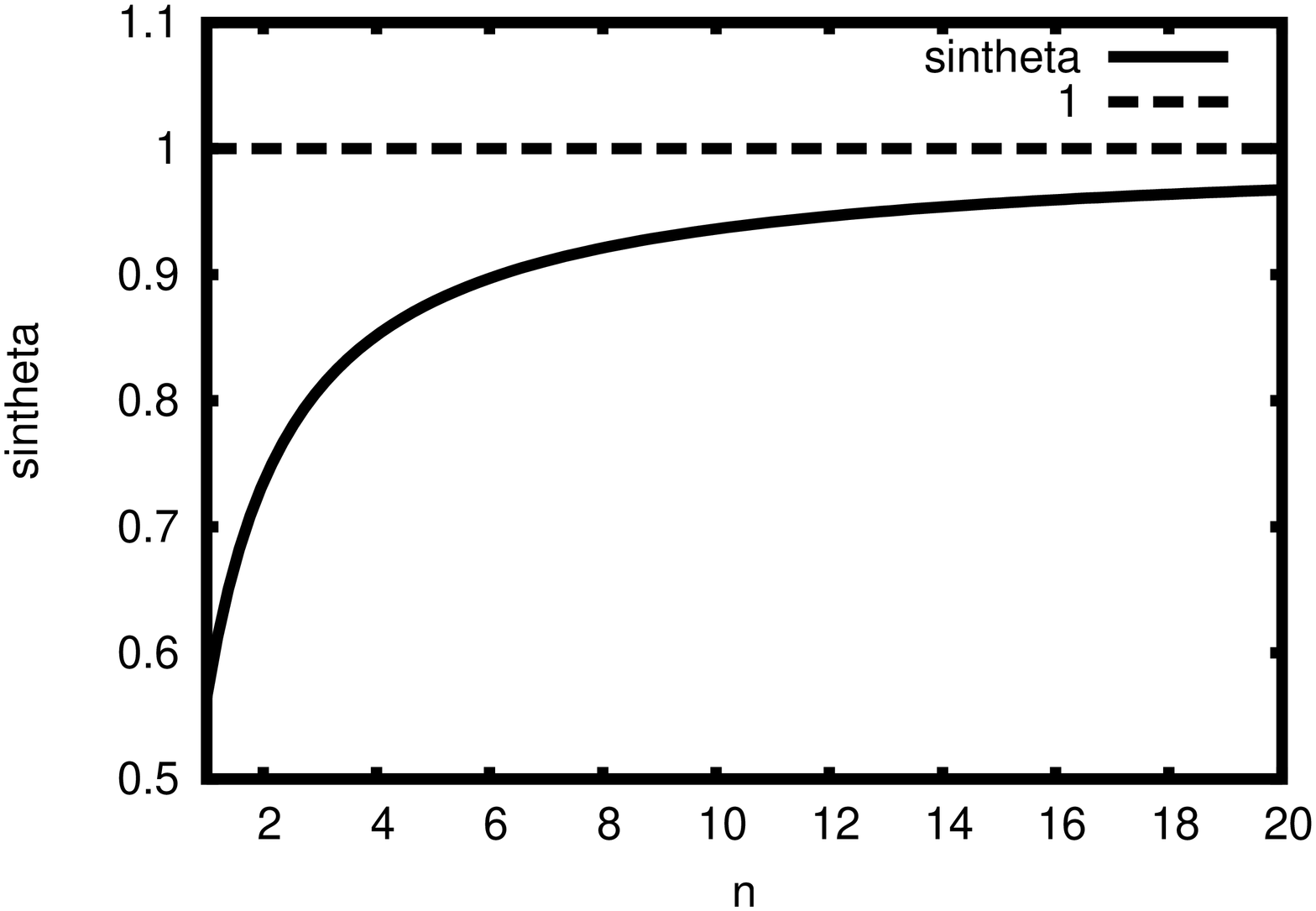}}}
\caption{$\sin 3 \theta$}
\label{fig:theta}
\end{minipage}
\begin{minipage}{80mm}
\rotatebox{270}{\resizebox{55mm}{!}{ \includegraphics{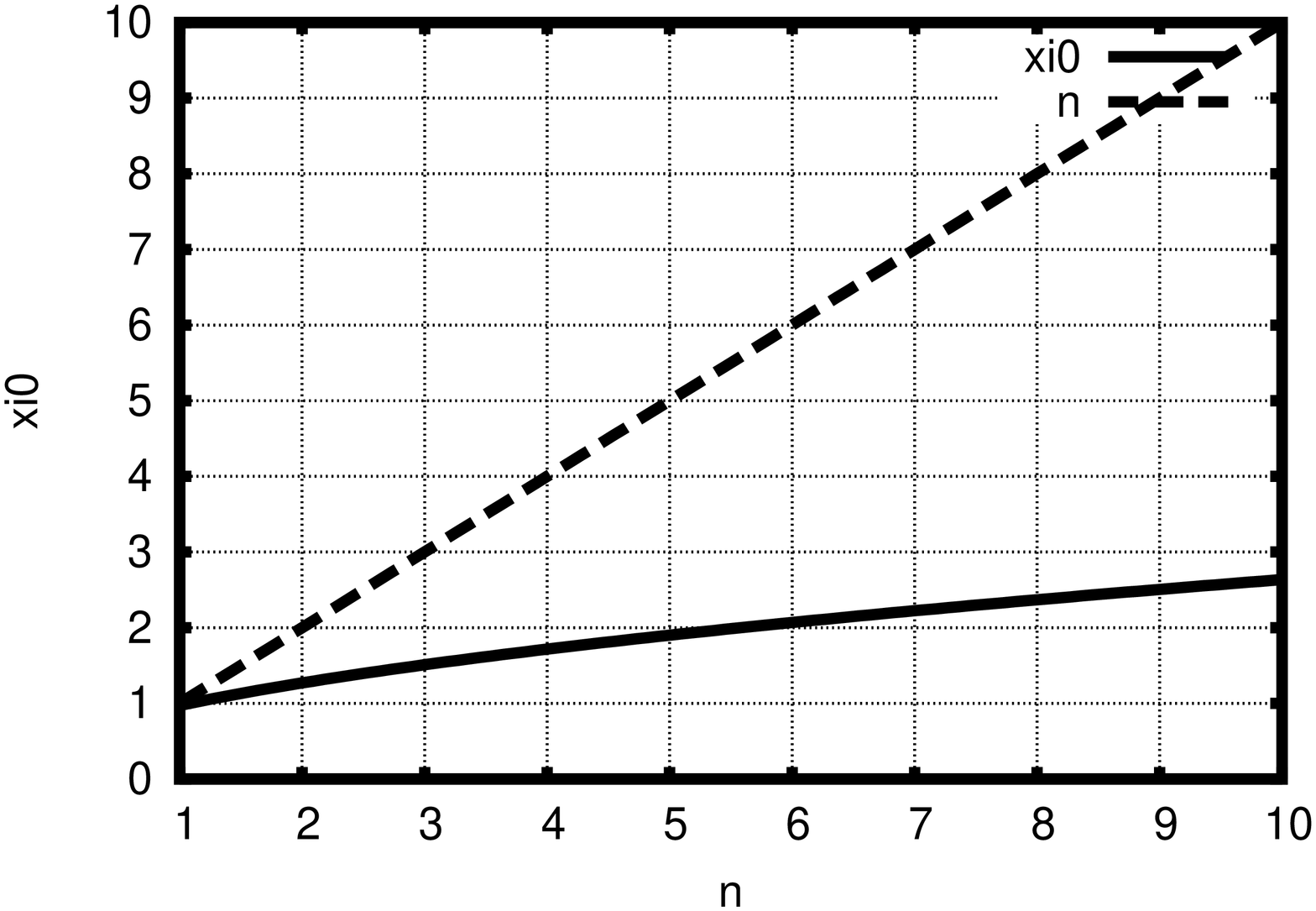}}}
\caption{$\xi_0$}
\label{fig:xi0}
\end{minipage}
 \end{figure}

Let us discuss the property of solutions. 
The condition $p > q$ implies that $2p > n$. Therefore $A_p,B_p$ and $C_p$  are positive. In this section suppose that $ [n/2]+1 \leq p \leq n$. 
Boundary conditions for the function $w(x)= y(x)/x(1-x)$ are as follows. At $\tau = + \infty$, $ \dot{f}/f = r f'/f \sim r^{2(p-q)-1} \rightarrow \infty$. Therefore $(x,w) \rightarrow ( 0,\infty)$. 
At $\tau = - \infty$, $\dot{f}/(1-f) = r f'/(1-f) \rightarrow - \lambda_1 $. $(x,w) = (1, -\lambda_1)$, where $\lambda_1 \geq 2- 1/(p-q)$. 
Until this point, $\lambda_!$ was a free parameter. Now we must impose that the point $(1,- \lambda_1)$ is a fixed point of the differential equation, 
\begin{align}
G_p(x,w,z) &= x(1-x)w z  - (2+A_p )x w^2 - B_p w  +C_p  ,& G_p\left( x,w, \frac{dw}{dx} \right)&=0 . 
\label{eq:abeldiffeq}
\end{align}
Fixed points of this equation are given as $(0, C_p/B_p)$,  $(1, w_{\pm} )$, where $w_{\pm}$ are defined as solutions of a quadratic equation, 
\begin{align}
(2+A_p ) w^2 +  B_p w  - C_p &=0 , &
w_{\pm} &= \frac{-B_p  \pm \sqrt{ B_p^2 + 4 C_p (2+A_p) }}{2(2+A_p)} ,& w_- &< 0 < w_+ .  
\end{align}
Therefore the exponent $\lambda_1$ is $-w_-$. 
In the case of $p=n$, $-w_-$ is 
\begin{align}
\lambda_1 = -w_- &= \frac{2n-1  +  (2n+1) }{2n} = 2. 
\end{align}
In the other cases, 
\begin{align}
\lambda_1 = -w_-  &= \frac{(4p-2n-1)  +  \sqrt{ (4p-2n-1)^2 + 4 (4n-4p+2) (3p-n) }}{2(3p-n)}  . 
\end{align}
Remember that $\lambda_1$ should satisfy the condition $\lambda_1 \leq 2 - 1/(p-q)$. 
\begin{align}
\lambda_1 - \left(  2 - \frac{1}{2p-n}  \right) &= \frac{(2p-n) \sqrt{\Delta} - (4p-n)(4p-2n-1)}{2(3p-n)(2p-n)},\cr
\Delta &=  (4p-2n-1)^2 + 4 (4n-4p+2) (3p-n)  .
\end{align}
The condition is interpreted as 
\begin{align}
&4 \xi^2 (n-\xi +1)    -  (\xi +n )  (2\xi-1)^2  \geq 0 , 
\end{align}
where $\xi := 2p-n$. 
Therefore we obtain a condition $\xi < \xi_1 , \xi_2 < \xi < \xi_0$, where $\xi_k (k=0,1,2)$ are defined  as
\begin{align}
\xi_k &= \frac{1}{3} + \sqrt{\frac{12n+5}{18}}  \cos \left( \theta + \frac{2\pi}{3} k \right) ,& \xi_1 &< \xi_2 < \frac{1}{3} <\xi_0 . 
\end{align}
Here the angle $\theta$ is defined as
\begin{align}
\theta &:=  \frac{1}{3} \arcsin \left( 6^3  \sqrt{ \frac{2\sigma}{(5+12n)^3  } }   \right),&
\sigma&:=  4\left( \frac{n}{6} + \frac{5}{72} \right)^3 - \left( \frac{n}{24} + \frac{7}{216} \right)^2,&
0&< \theta < \frac{\pi}{6} .
\end{align}
It turns out that $\sin 3 \theta$ increases monotonically as $n$ increases.  
And $\sin 3 \theta$ is larger than $0.55$ as is shown in Figure \ref{fig:theta}. Therefore $\xi_2 > 0$. 
We obtain, $1 \leq \xi \leq \xi_0$. $\xi_0$ is plotted in Figure \ref{fig:xi0}.  
Some examples are shown in Table \ref{tbl:example}. In the table, it can be read off that $n=4,p=3$ and $n=5, p=4$ do not satisfy the condition. 
\begin{table}
\begin{minipage}{60mm}
\begin{tabular}{ | l | l | l | l | l | l |}
\hline
$n$& $(p,q)$ & $\xi$ & $\lambda_1$ & $\eta_p$ & {\bf P} \\
\hline\hline
$1$ & $(1,0)$& 1 & $2$ & $1$ &$\bigcirc$ \\
\hline \hline
$2$ & $(2,0)$& 2 & $2$ & $3$&$\bigcirc$ \\ 
\hline\hline
$3$ & $(3,0)$ & 3 & $2$ & $5$&$\bigcirc$\\ \cline{2-5} 
& $(2,1)$ & 1 & $(1+\sqrt{73})/6$ & $1$ & $\bigcirc$\\ 
\hline
\end{tabular}
\end{minipage}
\begin{minipage}{50mm}
\begin{tabular}{ | l | l | l | l | l | l | }
\hline
$4$ & $(4,0)$ & 4 & $2$ & $7$ & $\bigcirc$\\ \cline{2-5} 
& $(3,1)$ & 2 & $(3+\sqrt{129})/10$ & $3$ &$\times$ \\ 
\hline\hline
$5$ & $(5,0)$ & 5 & $2$ & $9$ &$\bigcirc$ \\ \cline{2-5} 
& $(4,1)$  & 3 &  $ (5+\sqrt{193})/14$ & $5$ &$\times$ \\\cline{2-5} 
 & $(3,2)$  & 1 & $(1+\sqrt{161})/8$ & $1$ &$\bigcirc$\\
 \hline
\end{tabular}
\end{minipage}
\label{tbl:exponent}
\caption{Examples of $\lambda_1$.  The symbol {\bf P}  represents the property, $\lambda_1 > 2 - 1/(p-q)$. }
\label{tbl:example}
\end{table}

Let us consider the behavior of function $f$ around $\infty$. Remember that the exponent of $\ln f$ in (\ref{eq:behavioraroundinfinity}) is $\eta_p := 2(p-q)-1$. 
\begin{align}
f &= C \exp \left\{ - \rho_p x^{\eta_p} ( 1 + O(1/x))  \right\} , & x \rightarrow \infty .
\end{align}
In this region  $ x$ goes to  $0$ and $w$  diverges. Therefore the constant term $C_p$ in the differential equation (\ref{eq:abeldiffeq})  becomes relatively smaller than $B_p w$ and asymptotically the equation reduces to  
\begin{align}
 \frac{dw}{dx}   \sim  (2+A_p ) w +  \frac{B_p}{x}    . 
\end{align}
This is a linear equation and the solution is 
\begin{align}
w&\sim C(x) e^{(2+A_p) x} , & 
C'(x) & \sim \frac{B_p}{x}e^{-(2+A_p) x}   ,\cr
C(x) & = B_p {\rm Ei}( -(2+A_p) x ), & {\rm Ei} (x) &:=  - \int_{-x}^{\infty} \frac{e^{-t}}{t}dt  .
\end{align}
Here  ${\rm Ei}(x)$ is an exponential integral function. The function diverges as fast as logarithmic function in the limit $x \rightarrow 0$, 
\begin{align}
w &= {\boldsymbol C} + \ln (  (2+A_p) x ) + O(x \ln x) ,& {\boldsymbol C}&:= \lim_{N \rightarrow \infty} \left[ \sum_{k=1}^{N-1} \frac{1}{k} - \ln N  \right] , 
\end{align}
where ${\boldsymbol C}$ is Euler constant. 

\section{Summary}
In this article we consider generalized monopoles in SO($2n+1$) gauge theories on $(2n+1)$-dimensional space. 
We introduce new class of models and we consider Hedge-Hog type solutions in those models. 
Such a model is labeled by an integer $p$ which satisfies $[n/2]+1 \leq p \leq n$. 
Models consist of gauge field and scalar field which belongs to the vector representation ${\boldsymbol{2n+1}}$. 
The energy is defined as positive definite. 
We studied the topological property of the Bogomol'nyi bound in the sense of \cite{Arafune:1974uy}. 
We showed the topological current explicitly. 
In the case of $p=n, 1\leq 2p-n \leq \xi_0$, we showed that there can be Hedge-Hog solutions. 
Otherwise, the scalar field diverges when the distance $r$ from the monopole approaches to $0$. 
The divergence is harmful in order to consider the Prasad-Sommerfeld limit of the scalar potential term. 
The case $n=2,p=2$ is studied in \cite{Kihara:2004yz}. 

{\bf Acknowledgments : }
We are very grateful to Y. Hosotani and M. Nitta for their advise on this subject. 
We would like to thank S. Shimasaki for his comment.  
We are grateful to members in Korea Institute for Advanced Study for their support.



\begin{thebibliography}{99}

\bibitem{Dirac:1948um}
  P.~A.~M.~Dirac,
  Phys.\ Rev.\  {\bf 74} (1948) 817;
  T.~T.~Wu and C.~N.~Yang,
  Phys.\ Rev.\  D {\bf 12}, 3845 (1975);
  G.~'t Hooft,
  Nucl.\ Phys.\  B {\bf 79}, 276 (1974);
  A.~M.~Polyakov,
  JETP Lett.\  {\bf 20} (1974) 194
  [Pisma Zh.\ Eksp.\ Teor.\ Fiz.\  {\bf 20} (1974) 430];
  E.~B.~Bogomolny,
  Sov.\ J.\ Nucl.\ Phys.\  {\bf 24}, 449 (1976)
  [Yad.\ Fiz.\  {\bf 24}, 861 (1976)];
  M.~K.~Prasad and C.~M.~Sommerfield,
  Phys.\ Rev.\ Lett.\  {\bf 35}, 760 (1975);
  C.~N.~Yang,
  J.\ Math.\ Phys.\  {\bf 19}, 320 (1978).










\bibitem{Arafune:1974uy}
  J.~Arafune, P.~G.~O.~Freund and C.~J.~Goebel,
  J.\ Math.\ Phys.\  {\bf 16}, 433 (1975).


\bibitem{Tchrakian:1978sf}
  D.~H.~Tchrakian,
  J.\ Math.\ Phys.\  {\bf 21}, 166 (1980); 
  Phys.\ Lett.\  B {\bf 150}, 360 (1985);
  D.~H.~Tchrakian and F.~Zimmerschied,
  Phys.\ Rev.\  D {\bf 62}, 045002 (2000)
  [arXiv:hep-th/9912056];
  E.~Radu and D.~H.~Tchrakian,
  Phys.\ Rev.\  D {\bf 71}, 125013 (2005)
  [arXiv:hep-th/0502025].



\bibitem{Ito:1984wu}
  A.~Ito,
  Prog.\ Theor.\ Phys.\  {\bf 71} (1984) 1443;
  B.~Chen, H.~Itoyama and H.~Kihara,
  Mod.\ Phys.\ Lett.\  A {\bf 14}, 869 (1999)
  [arXiv:hep-th/9810237];
  B.~Chen, H.~Itoyama and H.~Kihara,
  Nucl.\ Phys.\  B {\bf 577}, 23 (2000)
  [arXiv:hep-th/9909075].
  T.~Tchrakian,
  Phys.\ Atom.\ Nucl.\  {\bf 71}, 1116 (2008).



\bibitem{Kihara:2004yz}
  H.~Kihara, Y.~Hosotani and M.~Nitta,
  Phys.\ Rev.\  D {\bf 71}, 041701 (2005)
  [arXiv:hep-th/0408068].
  H.~Kihara,
  Phys.\ Rev.\  D {\bf 77}, 127703 (2008)
  [arXiv:0802.3244 [hep-th]].










\bibitem{Arnold:1983}
V.~I.~Arnold,
 Springer (1983) ISBN~962-430-029-1. 








\end{thebibliography}
\end{document}